\documentclass[twoside,slac_one]{revtex4}
\usepackage{graphicx}
\usepackage{fancyhdr}
\usepackage{amsmath} 
\usepackage{bm}
\usepackage{amsxtra}
\usepackage{amssymb}
\usepackage{amsthm}
\usepackage{latexsym}
\usepackage{lscape}

\newcommand{\be}{\begin{equation}}
\newcommand{\ee}{\end{equation}}
\newcommand{\bp}{\begin{pmatrix}}
\newcommand{\ep}{\end{pmatrix}}
\newcommand{\bea}{\begin{eqnarray}}
\newcommand{\eea}{\end{eqnarray}}

\begin{document}

\title{Lorentz noninvariant neutrino oscillations without neutrino mass}

%

\author{K. Whisnant}
\affiliation{Department of Physics and Astronomy, Iowa State University, Ames, 
IA 50011, USA}

\begin{abstract}
The bicycle model of Lorentz noninvariant neutrino oscillations
without neutrino masses naturally predicts maximal mixing and a $1/E$
dependence of the oscillation argument for $\nu_\mu \to \nu_\tau$
oscillations of atmospheric and long-baseline neutrinos, but cannot
also simultaneously fit the data for solar neutrinos and KamLAND. We
examine all nineteen possible structures of the Standard
Model Extension for Lorentz noninvariant oscillations of massless
neutrinos that naturally have a $1/E$ dependence at high neutrino
energy. Due to the lack of any evidence for direction dependence, we
consider only direction-independent oscillations. Although we find a
number of models with a $1/E$ dependence for atmospheric and
long-baseline neutrinos, none can also simultaneously fit solar and
KamLAND data.
\end{abstract}

\maketitle

\thispagestyle{fancy}

\section{Introduction}

Neutrino data from atmospheric, long-baseline, solar and reactor
experiments are easily explained by oscillations of three active,
massive neutrinos~\cite{review}. Lorentz-invariance and $CPT$
violating interactions originating at the Planck scale can also lead
to neutrino oscillations. The Standard Model Extension
(SME)~\cite{SME} includes all such interactions that may arise from
spontaneous symmetry breaking but still preserve Standard Model gauge
invariance and power-counting renormalizability. Studies of neutrino
oscillations with Lorentz invariance violation have been made both for
massive~\cite{Coleman, BPWW, tandem} and massless~\cite{K1, K2,
bmw-bike} neutrinos. A model with nonrenormalizable Lorentz invariance
violating interactions and neutrino mass has also been
proposed~\cite{Diaz}.  However, no viable model has been found that
does not require at least one nonzero neutrino mass. The purpose of
this paper is to determine if Lorentz invariance violation alone can
account for the verified oscillation phenomena seen in atmospheric,
long-baseline, solar and reactor neutrinos. We do not attempt to fit
the possible oscillation signals seen in the LSND~\cite{LSND} and
MiniBooNE~\cite{miniboone} experiments. Our complete results are given
in Ref.~\cite{blmw}.

In the SME, the evolution of massless neutrinos in vacuum may be
described by the effective Hamiltonian~\cite{K1}
\be
(h_{eff})_{ij} = E \delta_{ij} + \frac{1}{E} \left[ a_L^\mu p_\mu
- c_L^{\mu\nu} p_\mu p_\nu \right]_{ij} \,,
\label{eq:heff}
\ee
where $p_\mu = (E, -E\hat p)$ is the neutrino four-momentum, $\hat p$
is the neutrino direction, $i,j$ are flavor indices, and $a_L \to -a_L$
for antineutrinos. The coefficients $a_L$ have dimensions of energy
and the $c_L$ are dimensionless. Direction dependence of the neutrino
evolution enters via the space components of $a_L$ and $c_L$, $\mu$ or $\nu =
X, Y, Z$, while direction independent terms have $\mu = \nu = T$. The
Kronecker delta term on the right-hand side of Eq.~(\ref{eq:heff}) may
be ignored since oscillations are insensitive to terms in $h_{eff}$
proportional to the identity.

The two-parameter bicycle model~\cite{K1} can be defined as follows:
$(c_L)_{ij}$ has only one nonzero element in flavor space and the only
nonzero $(a_L)_{ij}$ are $(a_L)_{e\mu} = (a_L)_{e\tau}$. These
interactions can be nonisotropic, which could lead to different
oscillation parameters for neutrinos propagating in different
directions. In Ref.~\cite{bmw-bike} it was shown that the pure
direction-dependent bicycle model is ruled out by solar neutrino data
alone, while a combination of atmospheric, solar and long-baseline
neutrino data excludes the pure direction-independent case.  A mixture
of direction-dependent and direction-independent terms (with 5 parameters) 
is also excluded when KamLAND data are added~\cite{bmw-bike}.

The key feature of the bicycle model is that even though the terms in
$h_{eff}$ are either constant or proportional to neutrino energy, at
high neutrino energies there is a seesaw type mechanism that leads to
$1/E$ behavior for the oscillation argument for atmospheric and
long-baseline neutrinos. In this paper we examine the general case of
direction-independent Lorentz invariance violation in the Standard
Model Extension for three neutrinos without neutrino mass, {\it i.e.},
Eq.~(\ref{eq:heff}) with only $c_L^{TT}$ and $a_L^T$ terms. We do not
consider possible direction-dependent terms since there is no evidence
for direction dependence in neutrino oscillation experiments (see,
{\it e.g.}, the experiments in Ref.~\cite{auerbach} and the analysis of
Ref.~\cite{K1}). For notational simplicity we
henceforth drop the $L$ subscript and $T$ superscripts from the
$c_L^{TT}$ and $a_L^T$ in our formulae.

We first look for textures of the $c_{ij}$ in flavor space that allow
a $1/E$ dependence of the oscillation argument at high neutrino
energy.  We then check the phenomenology for atmopheric,
long-baseline, solar and reactor neutrino experiments. We were unable
to find any texture of $h_{eff}$ that could simultaneously fit all
the data.

In Sec.~2 we review the constraints on the direction-independent
bicycle model. In Sec.~3 we list all possible textures of the $c$
coefficients and find which ones allow a $1/E$ dependence of the
oscillation argument at high neutrino energies. For those that do, we
first check the oscillation amplitude for atmospheric and
long-baseline neutrinos, and if suitable parameters are found we then
check the ability of the model to fit KamLAND and solar neutrino
data. In Sec.~4 we summarize our results.

\section{Neutrino oscillations in the bicycle model}

Neutrino oscillations occur due to eigenenergy differences in $h_{eff}$ and
the fact that the neutrino flavor eigenstates are not eigenstates of
$h_{eff}$. In our generalization of the direction-independent bicycle model,
\be
h_{eff} = \bp
-2cE + 2 a_{ee} & a_{e\mu} & a_{e\tau} \cr
a_{e\mu} & 0 & 0 \cr
a_{e\tau} & 0 & 0
\ep \,,
\ee
where the $c$ term is $CPT$-even and the $a_{ij}$ terms are $CPT$-odd.
The simple two-parameter bicycle model~\cite{K1} has $a_{e\tau} = a_{e\mu}$
and $a_{ee} = 0$. We allow $a_{e\mu}$ to be different from $a_{e\tau}$ so
that mixing of atmospheric neutrinos may be (slightly) nonmaximal. The
$a_{ee}$ term allows an adjustment of the oscillation probabilities of
low-energy solar neutrinos~\cite{K1}.

For large $E$, appropriate for atmospheric and long-baseline neutrinos, 
if $a^2 \equiv a_{e\mu}^2 + a_{e\tau}^2 \ll (cE)^2$, then the only appreciable
oscillation is
\be
P(\nu_\mu \leftrightarrow \nu_\tau) \simeq
\sin^22\phi \sin^2(\Delta_{32}L/2) \,,
\label{eq:Pmt}
\ee
where $\Delta_{32} \simeq a^2/(2cE)$ and $\tan\phi = a_{e\mu}/a_{e\tau}$.
The energy dependence of the oscillation argument in this limit is the same
as for conventional neutrino oscillations due to neutrino mass differences,
with an effective mass-squared difference $\delta m^2_{eff} = a^2/c$.
The measured value for $\delta m^2_{eff}$ in atmospheric and
long-baseline experiments then places a constraint that relates $a$ and $c$.

If $E$ is not too large, then the approximation in Eq.~\ref{eq:Pmt}
does not apply. Furthermore, in matter there is an additional term due
to coherent forward scattering~\cite{matter}, which adds a $\sqrt2 G_F
N_e$ term to the $\nu_e$-$\nu_e$ element of $h_{eff}$, where $N_e$ is
the electron number density. For adiabatic propagation in the sun, the
fact that the $^8$B neutrinos have an oscillation minimum $P_{min}
\simeq 0.30$ fixes $a$ to be $b \sqrt{2P_{min}/(1-2P_{min})} =
2.1\times10^{-12}$~eV, where $b \equiv G_F N_e^0/(2\sqrt2) =
1.7\times10^{-12}$~eV.

At very low energies the solar neutrino
oscillation probability is $P \approx 1 - \frac{1}{2}\sin^22\theta_{12}
\approx 0.57$, where $\theta_{12}$ is the usual solar neutrino mixing
angle~\cite{global-fits}. This gives two possible values for $a_{ee}$,
either $0.20 b$ or $-2.2 b$.

Finally, the $^8$B probability reaches
the minimum at $E_{min} = (a_{ee}+b)/c$, which must occur in the
energy region of the $^8$B solar neutrinos ($E_{min} \approx 10$~MeV),
which fixes the magnitude of $c$ to be $|c| = (a_{ee} + b)/E_{min} =
1.2~b/E_{min} \approx 2.0\times10^{-19}$.

We may now calculate the value of the atmospheric $\delta m^2_{eff}$
inferred from solar neutrino data: $\delta m^2 = a^2/c =
2.2\times10^{-5}$~eV$^2$, which is two orders of magnitude below the
measured value. Therefore the direction-independent bicycle model is
excluded.

\section{Other textures for $h_{eff}$}

\subsection{Classification of models}

There are six possible $c$ coefficients in $h_{eff}$: three real
diagonal coefficients and three complex off-diagonal coefficients (the
remaining three off-diagonals are fixed by the hermiticity of
$h_{eff}$). Therefore there are $2^6 = 64$ possible $c$ textures for
$h_{eff}$. Since the high-energy behavior of $h_{eff}$ is determined
by the $c$ coefficients, we classify the models by the number of
nonzero $c$ there are in $h_{eff}$. Within each main class there are
distinct subclasses which depend on the diagonal/off-diagonal
structure; within each subclass there are textures that differ only by
permutation of the flavor indices. In all there are 19
subclasses, which are listed in Table~\ref{table}.

\begin{table}[ht]
\begin{center}
\caption[]{A list of the 64 possible $c$ textures for $h_{eff}$.
The number in the subclass name corresponds to the number of nonzero
$c$, while the letters indicate a distinct diagonal/off-diagonal
structure (up to flavor permutation), if applicable. A $D_i$ in the
structure column indicates that a diagonal $c_{ii}$ is nonzero, while an
$O_{jk}$ indicates that off-diagonal $c_{jk}$ is nonzero. Different latin
indices in each case are distinct, {\it e.g.}, in the structure $D_i O_{jk}$
the diagonal element does not share a row or column with the
off-diagonal element, whereas for $D_{i} O_{ij}$ it does.}
\begin{tabular}{c l l c}\\
\hline
Number of & Subclass & Structure & Number of flavor\\
nonzero $c$ & & & permutations\\
\hline\hline
0 &  0 & $-$      & 1\\
\hline
1 & 1A & $D_i$    & 3\\
  & 1B & $O_{ij}$ & 3\\
\hline
2 & 2A & $D_i D_j$       & 3\\
  & 2B & $D_i O_{ij}$    & 6\\
  & 2C & $D_i O_{jk}$    & 3\\
  & 2D & $O_{ij} O_{ik}$ & 3\\
\hline
3 & 3A & $D_i D_j D_k$          & 1\\
  & 3B & $D_i D_j O_{ij}$       & 3\\
  & 3C & $D_i D_j O_{ik}$       & 6\\
  & 3D & $D_i O_{ij} O_{ik}$    & 3\\
  & 3E & $D_j O_{ij} O_{ik}$    & 6\\
  & 3F & $O_{ij} O_{ik} O_{jk}$ & 1\\
\hline
4 & 4A & $D_i D_j D_k O_{ij}$       & 3\\
  & 4B & $D_i D_j O_{ij} O_{ik}$    & 6\\
  & 4C & $D_i D_j O_{ik} O_{jk}$    & 3\\
  & 4D & $D_i O_{ij} O_{ik} O_{jk}$ & 3\\
\hline
5 & 5A & $D_i D_j D_k O_{ij} O_{ik}$    & 3\\
  & 5B & $D_i D_j O_{ij} O_{ik} O_{jk}$ & 3\\
\hline
6 &  6 & $D_i D_j D_k O_{ij} O_{ik} O_{jk}$ & 1\\
\hline\hline\\
\end{tabular}
\label{table}
\end{center}
\end{table}

We note that we may subtract any quantity proportional to the identity
from $h_{eff}$, since common phases in the neutrino equations of
motion do not affect the oscillations. In this way a diagonal element
may be removed or moved from one position to another. Then it is
not hard to see that the following subclasses are strictly equivalent:
3A$\leftrightarrow$2A, 3C$\leftrightarrow$3B, 4A$\leftrightarrow$3B,
4C$\leftrightarrow$4B, 5A$\leftrightarrow$4B and 6$\leftrightarrow$5B.

\subsection{Method for analyzing textures}

Our analysis proceeds as follows. We assume that $|c_{ij} E| \gg
|a_{k\ell}|$ for any $(i,j,k,\ell)$ for the high energies of
atmospheric and long-baseline neutrinos. This assumption is justified
since if any $a$ is similar in magnitude to the $cE$ at high
energies, then at lower energies (such as for reactor neutrinos) the
$a$ terms will dominate and the oscillation arguments will be
energy-independent, contrary to the KamLAND data, which measured a
spectral distortion (similarly, solar neutrinos would also not have an
energy-dependent oscillation probability, as they must).  Furthermore,
for the sake of naturalness, we assume that the nonzero $c$ coefficients
are all the same order of magnitude, and that likewise the nonzero $a$
coefficients are also the same order of magnitude.

Although for each texture the number of nonzero $c$ is determined,
initially we place no restrictions on the $a$. We note that if all
off-diagonal $c$ are nonzero, then by a redefinition of neutrino
phases and adding a term proportional to the identity we may take all
off-diagonal $c$ to be real and positive, except for one off-diagonal
$c$ that is complex (which we take to be $c_{e\tau}$ unless
otherwise noted). If any off-diagonal $c$ is zero, the nonzero
off-diagonal $c$ may all be taken as real and positive.

A key feature of the bicycle model was that even though the terms in
the effective Hamiltonian were either proportional to energy or
constant in energy, one eigenvalue difference was proportional to
$E^{-1}$, which mimics the energy dependence of the oscillations of
atmospheric and long-baseline neutrinos. Having an eigenvalue
difference proportional to $E^{-1}$ means that if the eigenvalues are
expanded in a power series in neutrino energy,
\bea
\lambda_i = \sum_{j=0}^\infty a_{ij} E^{1-j} \,,\qquad {\rm~for~}i=1,2,3 \,
\label{eq:lambda}
\eea
then two eigenvalues must be degenerate at leading order in $E$
(linear in $E$), and at the next order in energy ($E^0$, independent
of energy). Therefore in our analysis of more general three-neutrino
models with Lorentz invariance violation, we look for model parameters
that satisfy these conditions. Since an $L/E$ dependence has been seen
over many orders of magnitude in neutrino energy~\cite{superK}, it
seems likely that this is the only way the Hamiltonian in
Eq.~(\ref{eq:heff}) will be able to fit all atmospheric and
long-baseline neutrino data.

For each texture we expand the eigenvalues of $h_{eff}$
in powers of $E$ (as in Eq.~\ref{eq:lambda}),
where the leading $E^1$ behavior comes from the dominant $c E$
terms.  Since we want $1/E$ behavior for at least one oscillation
argument, we require that two of the eigenvalues be degenerate to
order $E^0$, with the first nonzero difference occurring at order
$E^{-1}$. In all cases this requirement puts constraints on the $c$
and $a$ coefficients. In our calculations we first find the
eigenvalues to order $E^1$ and impose the constraint that two
eigenvalues must be degenerate; then we find the eigenvalues of the
simplified $h_{eff}$ to order $E^0$ and again impose the degeneracy
condition. In this way the expressions for the eigenvalues to order
$E^{-1}$ will be made as simple as possible at each stage of the
calculation.

If the appropriate $1/E$ behavior can be achieved, the mixing angles
are then calculated to determine if $\nu_\mu$'s have maximal mixing
and $\nu_e$ small mixing for atmospheric and long-baseline
neutrinos. If the model is still viable, the energy dependences of
the oscillations of solar and KamLAND neutrinos are then 
checked for consistency.

At any time we are allowed to subtract a constant times the identity
matrix from $h_{eff}$, so that some classes of models are equivalent
to others.

\subsection{Models with $1/E$ behavior at high energies}

Only Classes 1A, 2C, 3B, 3F, 4D and 5B allow $1/E$ dependence of an
eigenvalue difference at high neutrino energies. As an example of a
class that does not have $1/E$ behavior, the effective Hamiltonian in
Class 1B is
\bea
h_{eff} = \bp
a_{ee} & cE + a_{e\mu} & a_{e\tau} \cr
cE + a_{e\mu}^* & a_{\mu\mu} & a_{\mu\tau} \cr
a_{e\tau}^* & a_{\mu\tau}^* & a_{\tau\tau}
\ep \,,
\eea
where $c_{e\mu} \equiv c$ may be taken as real and positive. The
eigenvalues to order $E^1$ are then $\lambda_1, \lambda_2 = \pm|c|E$
and $\lambda_3 = 0$.  Since these are all different at leading order
in $E$, they cannot give an oscillation argument proportional to
$E^{-1}$ at high energies, and this case is not allowed.

For Class 1A the effecive Hamiltonian is
\bea
h_{eff} = \bp
cE + a_{ee} & a_{e\mu} & a_{e\tau} \cr
a_{e\mu}^* & a_{\mu\mu} & a_{\mu\tau} \cr
a_{e\tau}^* & a_{\mu\tau}^* & a_{\tau\tau}
\ep \,,
\eea
where $c_{ee} \equiv c$ may be taken as real and positive. The
eigenvalues to order $E^0$ are then $\lambda_1 = cE + a_{ee}$ and
$\lambda_2,\lambda_3 = \frac{1}{2}[a_{\mu\mu} + a_{\tau\tau} \pm
\sqrt{(a_{\mu\mu}-a_{\tau\tau})^2 + 4|a_{\mu\tau}|^2}]$.  The
difference $\lambda_2 - \lambda_3$ can only be made zero to order
$E^0$ if $a_{\mu\mu} = a_{\tau\tau}$ and $|a_{\mu\tau}| = 0$. Then
$a_{\tau\tau}$ times the identity may be subtracted from $h_{eff}$; if
$a_{ee}-a_{\tau\tau}$ is redefined as $a_{ee}$, this case reduces to
the generalized bicycle model described in Sec.~2, which is excluded
by the combined data.

For Class 2C we have
\bea
h_{eff} = \bp
a_{ee} & a_{e\mu} & c_{e\tau}E + a_{e\tau} \cr
a_{e\mu}^* & c_{\mu\mu}E & a_{\mu\tau} \cr
c_{e\tau}E + a_{e\tau}^* & a_{\mu\tau}^* & a_{\tau\tau}
\ep \,,
\eea
where $c_{\mu\mu}$ and $c_{e\tau}$ may be taken as real and positive,
and we have subtracted a term proportional to the identity so that
$a_{\mu\mu} = 0$. The eigenvalues at leading order are $\lambda_1,
\lambda_2 = \mp c_{e\tau} E$ and $\lambda_3 = c_{\mu\mu}
E$. Degeneracy requires $c_{e\tau} = c_{\mu\mu}$. The oscillation
probabilities are approximately given by
\bea
P(\nu_\mu \to \nu_\mu) &=&
1 - \sin^22\theta \sin^2\left(\frac{\delta m^2_{eff}L}{4E}\right) \,,
\\
P(\nu_\mu \to \nu_e) &=& P(\nu_\mu \to \nu_\tau) =
\frac{1}{2}\sin^22\theta \sin^2\left(\frac{\delta m^2_{eff}L}{4E}\right) \,,
\eea
where $\sin\theta = |a_{ee} + a_{e\tau}|/\sqrt{2|a_{\mu\tau}|^2 + |a_{ee} +
a_{e\tau}|^2}$. Therefore a maximal oscillation amplitude for $\nu_\mu$ is
possible but $\nu_\mu$ oscillates equally to $\nu_e$ and $\nu_\tau$, which
is excluded by atmospheric neutrino experiments. Hence this case is not
allowed.

For Class 3F, the effective Hamiltonian is
\bea
h_{eff} = \bp
a_{ee} & c_{e\mu}E + a_{e\mu} & c_{e\tau}E + a_{e\tau} \cr
c_{e\mu}E + a_{e\mu}^* & a_{\mu\mu} & c_{\mu\tau}E + a_{\mu\tau} \cr
c_{e\tau}^*E + a_{e\tau}^* & c_{\mu\tau}E + a_{\mu\tau}^* & 0
\ep \,,
\eea
where $c_{e\mu}$ and $c_{\mu\tau}$ may be taken as real and positive,
$c_{e\tau}$ is complex and $a_{\tau\tau}$ has been set equal to
zero. In this case, for oscillations in atmospheric and long-baseline
neutrinos, all three flavors have oscillation probability
\begin{equation}
P(\nu_\alpha \rightarrow \nu_\alpha) = 
\frac{5}{9}-4 {|U_{\alpha 2}|}^2 \left(\frac{2}{3}-{|U_{\alpha 2}|}^2 \right)
\sin^2\left(\frac{\delta m^2_{eff} L}{4E} \right) \,,
\end{equation}
where $U$ is the matrix that diagonalizes $h_{eff}$.
This implies that all flavors of downward atmospheric neutrinos would
be suppressed by a factor of 5/9, which is contrary to the data.
Therefore this case is excluded.

Only three classes have the proper $1/E$ behavior {\it and}
oscillation amplitudes for high energy neutrinos: 3B, 4D and 5B. For
these cases we must check the predictions for solar and reactor neutrinos.
As an example, for Class 3B we have
\bea
h_{eff} = \bp
c_{ee}E + a_{ee} & a_{e\mu} & c_{e\tau}E + a_{e\tau} \cr
a_{e\mu}^* & 0 & a_{\mu\tau} \cr
c_{e\tau}E + a_{e\tau}^* & a_{\mu\tau}^* & c_{\tau\tau}E + a_{\tau\tau}
\ep \,,
\eea
where $c_{ee}$, $c_{\tau\tau}$ and $c_{e\tau}$ may be taken as real
and $a_{\mu\mu}$ has been set to zero by a subtraction proportional to
the identity. Degeneracy at order $E$ requires $c_{\tau\tau} = r
c_{e\tau} = r^2 c_{ee}$, where $r$ is a free parameter. Then the
oscillation probabilities for high-energy neutrinos are
\bea
P(\nu_\mu \to \nu_\mu) &=&
1 - \sin^22\theta \sin^2\left(\frac{\delta m^2_{eff}L}{4E} \right) \,,
\\
P(\nu_\mu \to \nu_e) &=&
\sin^2\phi \sin^22\theta \sin^2\left(\frac{\delta m^2_{eff}L}{4E} \right) \,,
\eea
where $\tan\phi = r$ and $\tan\theta = \sqrt{1+r^2}|a_{e\mu}|/|r
a_{ee}-a_{e\tau}|$.  Maximal $\nu_\mu$ oscillations are possible for
$\theta \simeq \pi/4$, which imposes the condition
$\sqrt{1+r^2}|a_{e\mu}| \simeq |ra_{ee} - a_{e\tau}|$.

Oscillations of $\nu_e$ at high energies must be small due to the
limit on $\nu_\mu \to \nu_e$ from K2K~\cite{K2K-nue} and
MINOS~\cite{MINOS-nue}. For K2K and MINOS the
oscillation amplitude for $P(\nu_\mu \to \nu_e)$, $\sin^2\phi
\sin^22\theta$, has an upper bound of about 0.14, which implies $r <
0.43$ for $\theta \simeq \pi/4$. The T2K experiment sees evidence for
$\nu_\mu \to \nu_e$ at the $2.5\sigma$ level~\cite{T2K}; the T2K
allowed regions are consistent with this bound.

\begin{figure}[ht]
\centering
\includegraphics[width=12cm]{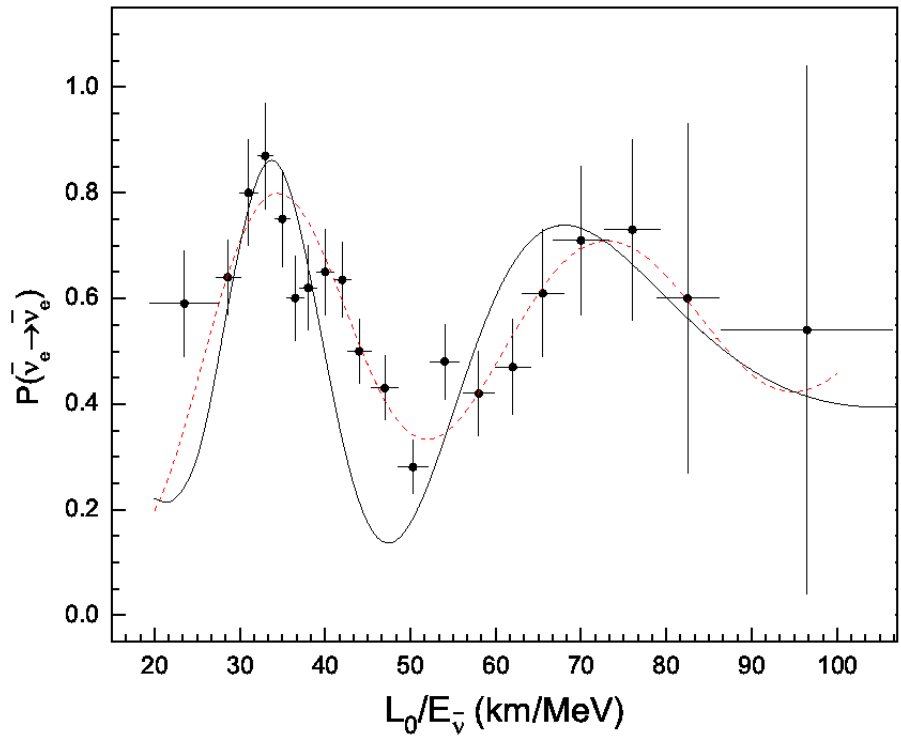}
\caption{The best fit to the KamLAND data for Class 3B (solid lines) and
the standard oscillation scenario with neutrino masses (dashed lines).}
\label{fig:3BKamLAND}
\end{figure}

For solar or reactor neutrinos the large energy limit does not apply.
Adjusting the remaining free parameters, KamLAND data can be fitted
reasonably well, as shown in Fig.~\ref{fig:3BKamLAND}. However,
the fit is not as good as the standard oscillation scenario
with neutrino mass.

\begin{figure}[!htb]
\centering
\includegraphics[width=12cm]{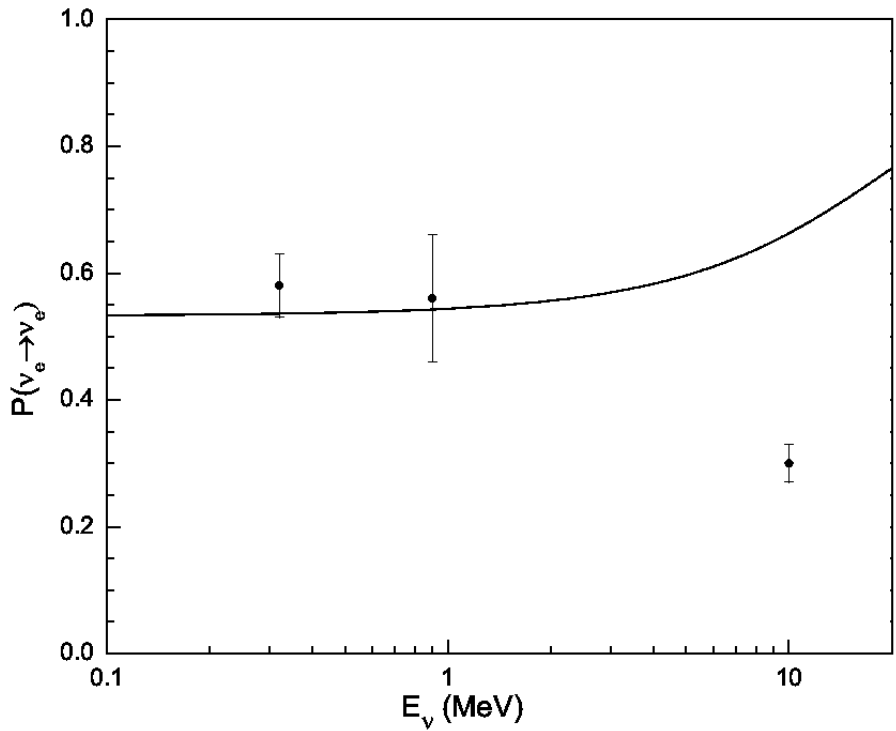}
\caption{The prediction of Class 3B for the oscillation probability
of solar neutrinos using the parameter values obtained from fitting
KamLAND data~\cite{kamland}. The solar data points are from
an update of the analysis in Ref.~\cite{BMW}.}
\label{fig:3Bsolar1}
\end{figure}

Next we use these parameter values to check the solar
phenomenology. Since the operator for \(a\) breaks $CPT$, we reverse
the sign of \(a\) when we apply these parameter values to the solar
neutrinos. However, the prediction does not agree with the solar data
at high energies given the upper bound on $r$ from above (see
Fig.~\ref{fig:3Bsolar1}).

\begin{figure}[!htb]
\centering
\includegraphics[width=12cm]{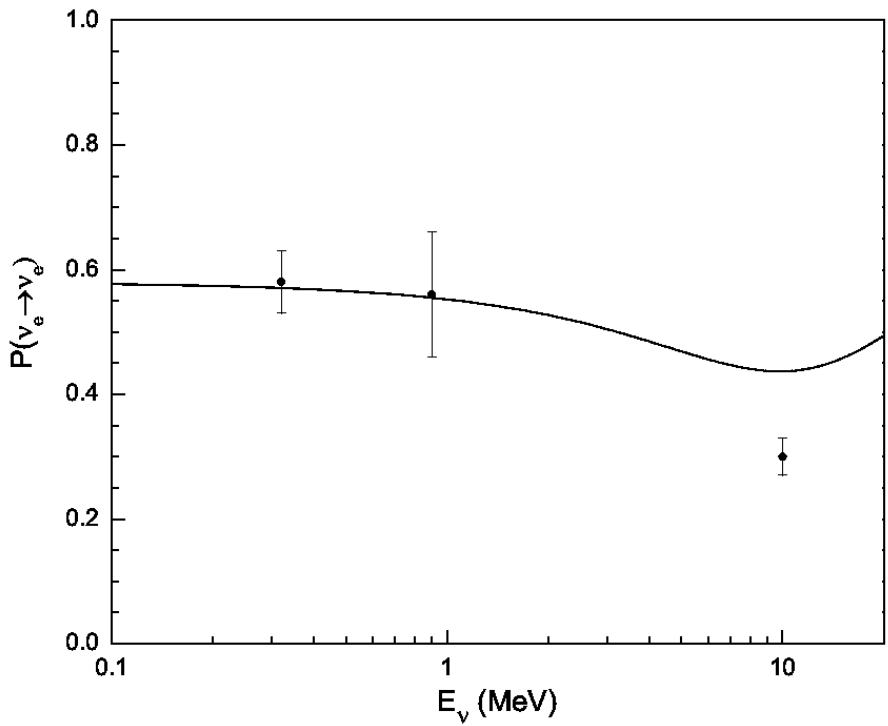}
\caption{Best fit prediction for survival probability of \(\nu_{e}\)
for solar neutrinos for Class 3B, assuming $r < 0.43$.}
\label{fig:3Bsolar2}
\end{figure}

We also searched the \(a_{ee}\), \(a_{e\tau}\) and \(r\) parameter
space to fit the solar data separately. The best fit still can not
yield reasonable agreement with the solar data at high energies for
$r < 0.43$ (see Fig.~\ref{fig:3Bsolar2}. If we do not impose the constraint
on \(r\), the fit to the solar data is improved at high energies. However,
we cannot simultaneously fit the KamLAND and solar data even with larger
\(r\). We found that we also need \(|a_{ee}|\) to become larger in
order to fit the solar data, but larger \(|a_{ee}|\) yields fast
oscillations for KamLAND data with averaged probabilities around
\(1/2\).

Classes 4D and 5B can also fit the KamLAND data, but not also solar data.
Therefore no case can fit all of the data simultaneously.

\section{Summary}

We have examined the general three neutrino effective Hamiltonian in
Eq.~(\ref{eq:heff}) for the case of direction-independent interactions
and no neutrino mass. We looked for texture classes in which two eigenvalues
were degenerate to order $1/E$ at high neutrino energy, so that
oscillations of atmospheric and long-baseline neutrinos would exhibit
the usual $L/E$ dependence.

Among the classes that had the proper $1/E$ dependence at high energy,
none was also able to fit the atmospheric, long-baseline, solar and
KamLAND data simultaneously. Class 1A (along with the equivalent Classes
2A and 3A) reduced to the direction-independent bicycle model, which has
been shown to be inconsistent with the solar, atmospheric and long-baseline
neutrino data. Classes 2C (and the equivalent 3E) and 3F did not have the
proper oscillation amplitudes for atmospheric neutrinos. Finally, Classes
3B (and the equivalent Classes 3C, 4A and 4D) and 5B (and the equivalent
Class 6) were able to fit atmospheric and long-baseline neutrino data, but
could not simultaneously fit KamLAND and solar data at lower neutrino
energies. The major difficulty in these latter classes was reproducing the
low survival probability of high-energy solar neutrinos.

Although we have not made an exhaustive search of the parameter space,
the fact that high-energy neutrinos exhibit an $L/E$ dependence in
their oscillations over many orders of magnitude in $E$ suggests that
the only way this can occur in the effective Hamiltonian described by
Eq.~(\ref{eq:heff}) is via the degeneracy of two eigenvalues to order
$1/E$. Since none of the cases where such a degeneracy occurs are also
able to fit all neutrino data simultaneously, it seems extremely
unlikely that any direction-independent SME model
without neutrino mass will provide a viable description of all
neutrino oscillation phenmomena. There is also strong evidence against
direction-dependent terms.  Furthermore, nonrenormalizable Lorentz
noninvariant effective Hamiltonians with higher powers of energy (as
in, {\it e.g.}, the model of Ref.~\cite{Diaz}) and no neutrino masses
would require additional degeneracy conditions. Therefore it appears
highly unlikely that Lorentz invariance violation alone can account for
all of the observed oscillation phenomena.

\section*{Acknowledgments}

We thank Wan-yu Ye for computational assistance in the early stages of
this work and A. Kostelecky for useful discussions. We also
thank the Aspen Center for Physics for its hospitality during the initial
stages of this work. This research was supported by the
U.S. Department of Energy under Grant Nos.~DE-FG02-95ER40896,
DE-FG02-01ER41155, and DE-FG02-04ER41308, by the NSF under Grant 
No. PHY-0544278, and by the Wisconsin Alumni Research
Foundation.

\end{document}